\documentclass[aps,pra,twocolumn,floatfix,reprint]{revtex4-1}
\usepackage{amsfonts}
\usepackage{mathrsfs}
\usepackage{amsmath}
\usepackage{color}
\usepackage{graphicx}
\usepackage{bm}
\usepackage{amssymb}
\usepackage{xspace}
\usepackage{epstopdf}
\usepackage{dcolumn}
\usepackage{tabularx}
\usepackage{longtable}
\usepackage[colorlinks=true, pdfstartview=FitV, linkcolor=blue, citecolor=blue, urlcolor=blue]{hyperref}
\usepackage[normalem]{ulem}

\begin{document}

\title{Expansion of harmonically trapped interacting particles and time
dependence of the contact}
\author{Chunlei Qu$^{1}$}
\email{chunleiqu@gmail.com}
\author{Lev P. Pitaevskii$^{1,2}$}
\author{Sandro Stringari$^{1}$}
\date{\today }

\affiliation{
$^{1}$INO-CNR BEC Center and Dipartimento di Fisica, Universit\`a di Trento, 38123 Povo, Italy \\
$^{2}$Kapitza Institute for Physical Problems RAS, Kosygina 2, 119334 Moscow, Russia}

\begin{abstract}
We study the expansion of an interacting atomic system at zero temperature,
following its release from an isotropic three-dimensional harmonic trap and
calculate the time dependence of its density and momentum distribution, with
special focus on the behavior of the contact parameter. We consider
different quantum systems, including the unitary Fermi gas of infinite
scattering length, the weakly interacting Bose gas, and two interacting
particles with highly asymmetric mass imbalance. In all cases analytic
results can be obtained, which show that the initial value of the contact,
fixing the $1/k^4$ tail of the momentum distribution, disappears for large
expansion times. Our results raise the problem of understanding the recent
experiment of Chang \textit{et al.} [Phys. Rev. Lett. \textbf{117}, 235303 (2016)] carried out on a
weakly interacting Bose gas of metastable $^4$He atoms, where a $1/r^4$ tail
in the density distribution was observed after a large expansion time,
implying the existence of the $1/k^4$ tail in the asymptotic momentum
distribution.
\end{abstract}

\maketitle

\section{Introduction}

It is well known that the momentum distribution $n(k)$ of a system of $N$
interacting particles \textit{normalized as $\int_0^\infty n(k) d^3k=N$}
exhibits, at equilibrium, the scaling behavior \textit{$(2\pi)^{-3}\mathcal{C%
}/k^4$} at large momenta $\hbar k$, where the coefficient $\mathcal{C}$ defines
the total Tan contact~\cite{Tan2008}, a universal quantity connecting the
thermodynamic properties of the many-body system to the short range behavior
of its wave function~\cite%
{Tan2008,Olshanii2003,Braaten2008,Zhang2009,Werner2009,Kuhnle2010,Barth2011}%
. In the case of a weakly interacting Bose gas this tail is physically
associated with the presence of the quantum depletion of the condensate, a
challenging feature well understood theoretically, but difficult to measure
experimentally~\cite{Wild2012,Xu2006,Utsunomiya2008}. The $1/k^4$ tail of
the momentum distribution has been measured in the unitary Fermi gas,
profiting from the possibility of turning off the interaction, just after the
release of the trap~\cite{Stewart2010,Sagi2012}. In this case the momentum
distribution is conserved during the expansion, so that experiments measure
the value of the contact before expansion. Despite the systematic
investigations of the contact at equilibrium~\cite%
{Haussmann1994,Palestini2010,Kuhnle2011,Drut2011,Hu2011,Valiente2011,Yan2013,Hoinka2013,Smith2014,Laurent2014}%
, its evolution in a dynamical context has been rarely explored~\cite%
{Makotyn2014,Sykes2014,Bardon2014,Zill2015,Corson2016,Fletcher2016}. The importance of this
question is manifested by the fact that the momentum distribution is
affected by interactions during the expansion and, for large expansion
times, when it becomes time-independent, can be directly related to the
measurable time-dependent density distribution $\rho(\mathbf{r},t)$ through
the ballistic relationship 
\begin{equation}
n_{\text{asym}}(\mathbf{k}) = \lim_{t \to \infty}(\hbar t/m)^3 \rho(\mathbf{r%
}=\hbar \mathbf{k}t/m, t),  \label{eq:ballistic}
\end{equation}
which assumes that, at large times, atoms move as free particles.
According to Eq.~(\ref{eq:ballistic}), an asymptotic $1/k^4$ tail in the
momentum distribution should show up in a tail of the density distribution,
characterized by the $t/r^4$ dependence for large values of $r$ and $t$. A
recent experiment~\cite{Chang2016}, where the density distribution of an
expanding gas of metastable $^4$He atoms was measured with high accuracy,
has indicated that the contact parameter $\mathcal{C}$ is actually increased
with respect to the initial value, and thus motivated us for the theoretical
investigation of the dynamical evolution of the contact.

The mechanism of the expansion of a dilute atomic gas is well understood in
the regime of distances satisfying the condition $r \le R_{\text{TF}}(t)$,
where $R_{\text{TF}}(t)$ is the Thomas-Fermi radius at the expansion time $t$%
, or momenta smaller than the inverse of the healing length~\cite%
{Pitaevskii2016}. In this regime approximate schemes based on scaling
transformations~\cite{Kagan1996,Castin1996} or on the hydrodynamic picture~%
\cite{Dalfovo1997} provide an accurate description of the expansion and, in
particular, have proven very successful in predicting the time dependence of
the aspect ratio, a quantity measurable with high precision.

The description of the expansion at a more microscopic level, involving the
large-$k$ components of the momentum distribution, and hence the
determination of the time dependence of the contact, represents a more
challenging problem, whose solution requires the implementation of
time-dependent dynamic theories beyond the mean field level. The purpose of
the present work is to discuss some examples where such theories can be
worked out explicitly and the question of the behavior of the tails of the
momentum distribution, when the interaction is not turned off during
the expansion, can be addressed in a quantitative way at zero temperature.
These include the unitary Fermi gas, the weakly interacting Bose gas and the
two-body problem with highly asymmetric mass imbalance which reduces the
problem to the solution of the single-particle Schr\"odinger equation.

The paper is structured as follows. In Sec.~\ref{sec:manybody} we
investigate the evolution of the contact for the unitary Fermi gas and the
weakly repulsive Bose gas. In Sec.~\ref{sec:twobody} we explicitly calculate
the density and momentum distributions during the expansion for the
two-particle system, both at unitarity and in the case of weakly repulsive
interaction. In both cases the contact is found to decrease for large
expansion times in the same way as in the corresponding many-body systems.
We also calculate the change of the contact after a sudden quench of the
scattering length from a small value to infinity just before the expansion.
Our concluding remarks are reported in Sec.~\ref{sec:conclusion}.

\section{Many-body interacting systems}

\label{sec:manybody}

\subsection{Unitary Fermi gas}

A remarkable feature exhibited by the Fermi gas at unitarity is that an
exact solution of the problem of the expansion from a three-dimensional (3D)
isotropic harmonic trap can be formulated in terms of the exact scaling
transformation~\cite{Castin2004} 
\begin{equation}
\psi(\mathbf{r}_1,\ldots,\mathbf{r}_N,t)=\mathcal{N}(t)e^{i\sum_{j}(
mr_j^2/2\hbar)(\dot{b}/b)}\psi_0\left(\frac{\mathbf{r}_1}{b},\ldots,%
\frac{\mathbf{r}_N}{b}\right)  \label{castin}
\end{equation}
of the many-body wave function. Here $N$ is the total particle
number of the two spin states, $|\mathcal{N}(t)|=b^{-3N/2}$ is the
normalization constant, and $b(\tau)=\sqrt{1+\tau^2}$ with $\tau=\omega_{%
\text{ho}}t$ the dimensionless time and $\omega_{\text{ho}}$ the trapping
frequency.

Consequently, the density distribution of the system during the expansion
is given by 
\begin{equation}
\rho(r,\tau)= \frac{1}{(1+\tau^2)^{3/2}}\rho_0(r/\sqrt{1+\tau^2}),
\label{castindensity}
\end{equation}
where $\rho_0(r)$ is the equilibrium density evaluated at $t=0$. The
ballistic relation [Eq.~(\ref{eq:ballistic})] then gives the following
expression for the asymptotic momentum distribution :
\begin{equation}
n_{\text{asym}}(k)= (\hbar/m\omega_{\text{ho}})^3 \rho_0(r= \hbar k/m\omega_{%
\text{ho}}),  \label{eq:asym_unitary}
\end{equation}
which holds for large times satisfying the condition $t \gg 1 /\omega_{\text{%
ho}}$. The above results [Eq.~(\ref{castin})-(\ref{eq:asym_unitary})] hold if the gas preserves the conditions of
unitarity (infinite scattering length) during the expansion. If the
scattering length is instead modified or switched off just after the release
of the trap, the asymptotic momentum distribution exhibits a different
behavior (see, for example, Refs.~\cite{Stewart2010,Sagi2012}). The scaling
result in Eq.~(\ref{eq:asym_unitary}) shows that the occurrence of the $%
1/k^4 $ tail in $n_{\text{asym}}(k)$ should be associated with a
corresponding $1/r^4$ tail at large $r$ in the equilibrium density
distribution before expansion. This behavior is ruled out by the
simple argument that the $1/r^4$ law would imply an unphysical infrared
divergent behavior in the ground state harmonic oscillator energy and hence,
due to the virial theorem, in the total ground state energy. The absence of
the $1/r^4$ can be explicitly shown in the case of the two-body problem (see
Sect. IIIA).

An explicit expression for the time dependence of the contact can be
obtained by making the adiabatic ansatz \cite{Tan2008}. This ansatz
corresponds to assuming that, during the expansion, the large momentum
component of the momentum distribution is given by the same expression $%
(2\pi)^{-3}\mathcal{C}(t)/k^4$ holding at equilibrium, with $%
\mathcal{C}(t)$ evaluated using the time-dependent value of the density
profile. The adiabatic ansatz holds if the two-body interaction, which
dictates the short range behavior of the wave function, is kept constant or
slowly varying in time. In the unitary Fermi gas the contact is given by 
$\mathcal{C}(t)=\int d\mathbf{r} \{\alpha [3\pi^2 \rho(\mathbf{r}%
,t)]^{4/3}\}$ where $\alpha\simeq 0.12$ is a universal parameter. Using the
position and time dependence of the density predicted by the scaling law
(Eq.~(\ref{castindensity})), the contact is then expected to evolve in time
as $\mathcal{C}(t)=\mathcal{C}_0/(1+\omega_{\text{ho}}^2 t^2)^{1/2}$ where 
$\mathcal{C}_0=256\pi\alpha N k_{\text{F}}^0/(35\xi_{\text{B}%
}^{1/4}) $ is the initial contact of the unitary Fermi gas at equilibrium, $%
\xi_\text{B}$ is the Bertsch parameter, and $k_{\text{F}}^0=[3\pi^2%
\rho_0(0)]^{1/3}$ is the initial local Fermi momentum at the trap center.

It is worth noticing that the same result [Eq.~(\ref%
{eq:asym_unitary})] for the asymptotic momentum distribution and for the
time dependence of the contact can be also derived by studying the behavior
of the one-body density matrix (see Appendix~\ref{appdenx1}),
thereby justifying the use of the adiabatic theorem.

\subsection{Weakly repulsive Bose gas}

Although in the case of the 3D Bose gas an exact scaling solution of
the many-body wave function similar to Eq.~(\ref{castin}) is not available,
the adiabatic ansatz discussed above is still expected to hold. In the local density
approximation, the
time dependence of the contact is expressed in the form $\mathcal{C}(t)=\int d%
\mathbf{r}[16\pi ^{2}\rho ^{2}(\mathbf{r},t)a^{2}]$ where $a$ is the $s$%
-wave scattering length and the integrand is the local contact density at
equilibrium predicted by Bogoliubov theory~(see \cite{Pitaevskii2016}, Sec.~18.3). 
We can then  use the time dependence of the density
profile during the expansion, as predicted by the scaling transformation $%
\rho (r,t)=\rho _{0}[r/b(t)]/b^{3}(t)$ of hydrodynamic theory, where $b(t)$ is the
relevant scaling parameter characterizing the expansion of a dilute Bose
gas~behaving, at large times, as $b(t)\propto \omega _{\text{ho}}t$ (see 
\cite{Castin1996,Kagan1996,Dalfovo1997} and \cite{Pitaevskii2016}, Sec.
12.7).  As a consequence, the contact of the dilute Bose gas decays as $%
\mathcal{C}(t)=\mathcal{C}_{0}/b^{3}(t)\propto \mathcal{C}_{0}/(\omega _{%
\text{ho}}t)^{3}$ at large times where $\mathcal{C}_{0}=120\pi N^{2}a^{2}/[7(R_{%
\text{TF}})^{3}]$ is the initial contact with $R_{\text{TF}}$, the initial
Thomas-Fermi radius of the trapped gas.

\section{Two-body problem}

\label{sec:twobody}

More complete and explicit results for the time evolution of the contact are
available by solving the expansion dynamics of a two-body problem, which
gives access to the exact time dependence of the density and momentum
distributions and provides explicit conditions for the applicability of the
adiabaticity ansatz. For the sake of brevity, we consider two interacting
particles in an isotropic 3D harmonic trap with one of them being so heavy
that the center-of-mass motion can be ignored. This reduces the problem to
the investigation of a one-body system described by the Hamiltonian 
\begin{equation}
H=-\frac{\hbar^2}{2m}\nabla^2+\frac{m}{2}\omega_{\text{ho}}^2r^2+V(r),
\label{eq:Hamiltonian}
\end{equation}
where $m$ is the mass of the light particle. $V(r)$ is the two-body
interaction characterized by the value of the $s$-wave scattering length, a
quantity which, in the presence of a Fano-Feshbach resonance, can be tuned
by applying an external magnetic field~\cite{Chin2010}. In this work, we
will consider the regularized pseudopotential $V(r)=(2\pi\hbar^2a/m)%
\delta^{(3)}(\mathbf{r})\frac{\partial}{\partial r}r$, whose eigenvalue
problem can be solved analytically~\cite{Busch1998}. The eigenfunctions take
the form 
\begin{equation}
\psi_0^{\mathbf{\nu}}(r)=\frac{A}{\sqrt{\pi}}\frac{a}{a_{\text{ho}}^{5/2}}%
e^{-r^2/2a_{\text{ho}}^2}\Gamma(-\nu)U\left(-\nu,\frac{3}{2},\frac{r^2}{a_{%
\text{ho}}^2}\right),  \label{eq:eigenfunction}
\end{equation}
where $U(x,y,z)$ is the hypergeometric function, $A$ is the normalization
constant, $a_{\text{ho}}=\sqrt{\hbar/m\omega_{\text{ho}}}$ is the harmonic
oscillator length, and $\nu=(E/2\hbar\omega_{\text{ho}}-3/4)$, with the
eigenvalue $E$ determined by 
\begin{equation}
\frac{\Gamma(-E/2\hbar\omega_{\text{ho}}+3/4)}{\Gamma(-E/2\hbar\omega_{\text{%
ho}}+1/4)}=\frac{a_{\text{ho}}}{2a}.  \label{eq:eigenvalue}
\end{equation}
Two different values of scattering length will be discussed in the following.

\subsection{Unitary interaction}

At unitarity ($a=\infty$), the ground state energy is $E=\hbar\omega_{\text{%
ho}}/2$ and the corresponding wave function takes the form $\psi_0(r)=\frac{1%
}{\sqrt{2}\pi^{3/4}}a_{\text{ho}}^{-1/2}\exp\left(-\frac{r^2}{2a_{\text{ho}%
}^2}\right)r^{-1}$. The wave function decays exponentially at large $r$
whereas it approaches $1/r$ for small $r$. We notice that the size of the
pair is reduced with respect to the non-interacting value according to $%
\langle r^2 \rangle_\infty / \langle r^2 \rangle _0=1/3$, reflecting the
attractive nature of the force at unitarity.

The ground state wave function in momentum space is readily obtained after a
Fourier transformation: 
\begin{equation}
\phi_0(k)=\sqrt{\frac{2}{\pi}}\int_0^\infty \psi_0(r)\frac{\sin(kr)}{kr}%
r^2dr = \frac{\sqrt{2}}{\pi^{5/4}}\frac{a_{\text{ho}}^{1/2}}{k}F\left(\frac{%
ka_{\text{ho}}}{\sqrt{2}}\right), \\
\end{equation}
where $F(z)=e^{-z^2}\int_0^{z}e^{y^2}dy$ is the Dawson function. Using its
asymptotic behavior~\cite{Dawson}, one finds the result 
\begin{equation}
n_0(k) =\frac{2a_{\text{ho}}}{\pi^{5/2}k^2}\left[F\left(\frac{ka_{\text{ho}}%
}{\sqrt{2}} \right) \right]^2 = \left\{%
\begin{array}{lll}
\pi^{-5/2}a_{\text{ho}}^{3}, & k\rightarrow 0 &  \\ 
\frac{\pi^{-5/2}a_{\text{ho}}^{-1}}{k^4}, & k\rightarrow \infty & 
\end{array}%
\right. \\
\end{equation}
for the momentum distribution, yielding the value \textit{$\mathcal{C}_0=8%
\sqrt{\pi}/a_\text{ho}$} for the contact.

When the particle is released from the isotropic trap, the wave function
evolves according to the exact scaling transformation, Eq.~(\ref{castin})
(with $N=1$), thus the expanding density distribution is given by 
\begin{equation}
\rho(r, \tau)=\frac{1}{2\pi^{3/2}(1+\tau^2)^{1/2}}\exp\left[-\frac{r^2}{%
(1+\tau^2)a_{\text{ho}}^2}\right]\frac{1}{r^2a_{\text{ho}}},
\label{eq:unitaryrhot}
\end{equation}
where $\tau=\omega_{\text{ho}} t$ is the dimensionless time. The density
distribution exponentially decays at large $r$ and the absence of an
additional $1/r^4$ tail indicates that the $1/k^4$ tail of the momentum
distribution will disappear at large expansion time. The time evolution of
the momentum distribution can be explicitly obtained by calculating the wave
function in momentum space. One finds 
\begin{eqnarray}  \label{phik}
\phi(k, \tau) =\left(\frac{1+i\tau}{1-i\tau} \right)^{1/4}\frac{\sqrt{2}}{%
\pi^{5/4}} \frac{a_{\text{ho}}^{1/2}}{k}F\left(\sqrt{1+i\tau}\frac{ka_{\text{%
ho}}}{\sqrt{2}} \right).  \notag \\
\end{eqnarray}
Therefore, as a consequence of the asymptotic behavior of the Dawson
function~\cite{Dawson}, we find that, for any fixed value of $t$, $n(k, t)=%
(2\pi)^{-3}\mathcal{C}(t)/k^4$ for $ka_\text{ho} \to \infty$ with the
contact $\mathcal{C}(t)=\mathcal{C}_0/\sqrt{1+\omega_{\text{ho}}^2t^2}$
vanishing like $1/t$ for large expansion times~\cite{footnote0}. This is the
same law obtained in the case of the unitary Fermi gas discussed in the
first part of the paper, using the adiabaticity ansatz for the evolution of
the contact.

\begin{figure}[tbp]
\includegraphics[width=7.5cm]{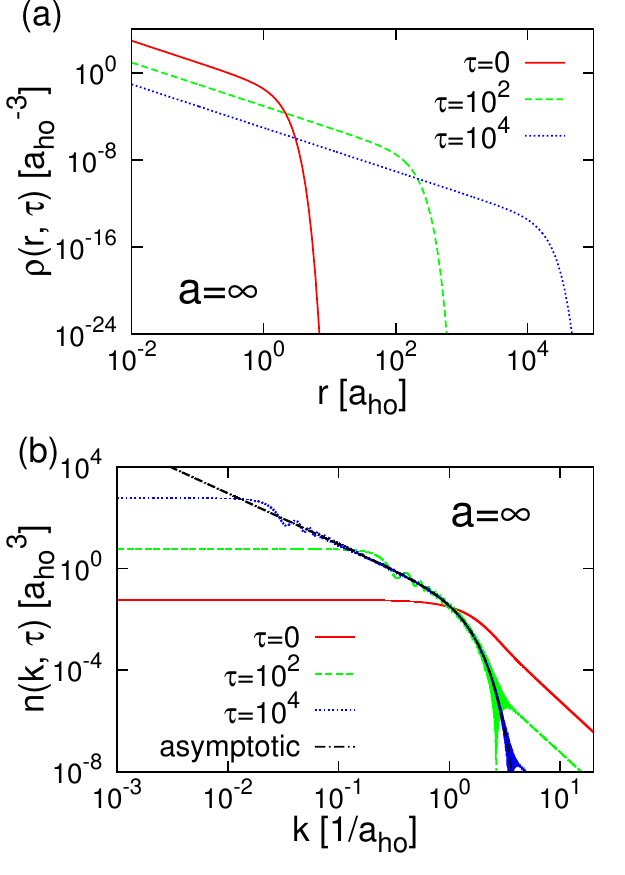}
\caption{(a) The density distribution $\protect\rho(r, 
\protect\tau)$ and (b) momentum distribution $n(k, \protect\tau)$ in the
presence of a regularized pseudopotential with unitary interaction applied
to the two-body problem. Different lines are the distributions at different
expansion time $\protect\tau=0$ (solid red), $10^2$ (dashed green), and $10^4$ (dotted blue), and the
asymptotic momentum distribution(dot-dashed black). Results are shown in
log-log scale plot. For finite expansion time, the density distribution $%
\protect\rho(r, \protect\tau)$ exhibits a $1/r^2$ behavior at small $r$ and
the momentum distribution $n(k, \protect\tau)$ exhibits a $1/k^4$ behavior
at large $k$.}
\label{fig:unitarity}
\end{figure}

The asymptotic momentum distribution, easily derivable by inserting Eq.(\ref%
{eq:unitaryrhot}) into the ballistic formula [Eq.(\ref{eq:ballistic})],
takes the form 
\begin{equation}
n_{\text{asym}}(k)=\frac{1}{2\pi^{3/2}} \frac{a_{\text{ho}}}{k^2}e^{-k^2a_{%
\text{ho}}^2}.  \label{eq:asym_unitary2body}
\end{equation}
The same expression for the asymptotic momentum distribution can be also
obtained using the asymptotic $F(z) \to i\sqrt{\pi} e^{-z^{2}}/2$ of the
Dawson function, holding for $\vert z \vert \to \infty$ and $\arg (z) \to
\pi/4$. This result can be obtained using the relation between Dawson
function and the error function $F(z)=i\sqrt{\pi}e^{-z^2}\text{erf}(-iz)/2$
and Eq. (7.2.4) from Ref.~\cite{handbook}.

Figure~\ref{fig:unitarity} shows the density distribution $\rho(r, \tau)$
and the momentum distribution $n(k, \tau)$ at $\tau=0, 10^2, 10^4$. The
figure shows that the value of $k$, above which one can observe the $1/k^4$
behavior, becomes larger and larger with the increase of $\tau$. We find that
for $\tau=10^4$ the $1/k^4$ tail is practically no longer visible and that
the momentum distribution [see Fig.~\ref{fig:unitarity}(b)], is
indistinguishable from the asymptotic behavior, except for small values of $%
k $ where reaching the asymptotic behavior requires even larger expansion
times. It is also worth noting that, for shorter times, where the $1/k^4$
tail is still visible in the momentum distribution, there is no trace of the 
$1/r^4$ tail in the density distribution [see Fig.~\ref{fig:unitarity}(a)]
because the ballistic relationship (\ref{eq:ballistic}) is not applicable on
these times.

\subsection{Weakly repulsive interaction}

In this case we assume that the system initially occupies the second
energy level of Eq.~(\ref{eq:eigenvalue}) to avoid the lowest energy level
of the pseudopotential interaction which corresponds, in the absence of the
trap, to a bound state if $a>0$. Note that, despite the absence of overlap
between the two states at $t=0$, the wave function may have a finite overlap
with the bound molecular state during the expansion, resulting in a
molecular contribution to the contact of the expanding configuration~\cite{Corson2015}. The
small probability of this process will be calculated at the end of the
section.

The wave function of the trapped state can be approximated, for $%
r\ll a_{\mathrm{ho}}$, by $\psi _{0}(r)\approx Aa_{\text{ho}}^{-3/2}(a/r-1)$%
, whereas its behavior at large $r$ is dominated by the exponential factor.
For the weak interaction case $a \ll a_{\mathrm{ho}}$ ($a=0.05a_\text{ho}$) the energy of the
state is $E\approx 3\hbar \omega _{\mathrm{ho}}/2$ and $A\approx 1/\pi
^{3/4}.$ The momentum distribution at large $k$ is determined by the small $%
r $ behavior of the wave function. One finds the dependence $n_{0}(k)=2A^2a^2/(\pi a_{\text{ho}}^{3}k^{4})$ for large $k$, yielding the value $%
\mathcal{C}_{0}=16\pi^2 A^2a^2/a_\text{ho}^3$ for the contact which is
proportional to $a^{2}$ like the case of the weakly interacting Bose gas.

The problem of the expansion can be solved in a way similar to that of
Ref.~\cite{Sykes2014}, where the short time evolution of the contact was
studied after quenching the scattering length to infinity in the presence of
a harmonic trap. After the release of the trap, the eigenfunctions of the
continuous spectrum with energy $E_q=\hbar^2q^2/2m$ are 
\begin{equation}  \label{basis}
R_q(r)=\frac{2a_{\text{ho}}^{-1/2}\sin(qr-\delta_q)}{r},
\end{equation}
where the phase shift $\delta_q$ should be determined by the Bethe-Peierls
boundary condition $\partial_r(rR_q)/(rR_q)|_{r\to 0}=-1/a$, and one finds $%
\tan\delta_q=qa$. At unitarity, $\delta_q=\pi/2$ and thus $R_q(r)=-2a_{\text{%
ho}}^{-1/2}\cos(qr)/r$. For ideal gas without interaction, $\delta_q=0$ and
thus $R_q(r)=2a_{\text{ho}}^{-1/2}\sin(qr)/r$. Furthermore, the functions $%
R_q$ satisfy the orthogonality condition $\int_{0}^\infty r^2 R_q(r)
R_{q^\prime}(r)dr=2\pi a_{\text{ho}}^{-1} \delta(q-q^\prime)$~\cite{Landau}.

The projection of the initial wave function $\psi_0(r)$ on such a basis
gives 
\begin{equation}
c(q)=\int_{0}^\infty R_q(r)\psi_0(r)r^2dr,  \label{eq:cp}
\end{equation}
which provides a useful quantity allowing for the calculation of the density
and momentum distributions as a function of time. The behavior of $c(q)$ at
large momenta satisfying the condition $q\gg a_{\text{ho}}^{-1}$ is related
to the behavior of the wave functions at short distances satisfying the
condition $r\ll a_{\text{ho}}$. Substituting the approximate wave function $%
Aa_{\text{ho}}^{-3/2}(a/r-1)$ into Eq.~(\ref{eq:cp}) and keeping terms up to
the first order in $a$, we find $c(q)\approx 2Aaa_{\text{ho}}^{-2}/q$ if the
interaction is turned off and $c(q)=0$ if the interaction is present during
the expansion. In the latter case, the \textit{approximate} initial wave
function $\psi_0$ is orthogonal to $R_q$ as it does not account for the
effect of the harmonic trap. A more accurate expression for $c(q)$ is
obtained by taking into account that the \textit{exact} initial wave
function is only approximately orthogonal to $R_q$ and satisfies the
equation $\left(-\frac{\hbar^2}{2m}\frac{d^2}{dr^2}+\frac{1}{2}m\omega_{%
\text{ho}}^2r^2-E\right)\left(r\psi_0\right)=0$. On the other hand, the wave
function $R_q(r)$ satisfies the equation $\left(-\frac{\hbar^2}{2m}\frac{d^2%
}{dr^2}-E_q \right)\left(rR_q \right)=0$. Multiplying the first equation by $%
rR_q(r)$ and the second one by $r\psi_0(r)$, followed by an integration with
respect to $r$, and using the fact that \textit{both} wave functions satisfy
the Bethe-Pierels boundary condition at $r=0$, we get 
\begin{equation}  \label{eq:cp2}
c(q)=\frac{\int_{0}^\infty (m\omega_{\text{ho}}^2r^2/2)(rR_q)(r\psi_0)dr}{%
E-\hbar^2q^2/2m}.
\end{equation}
Note that at the value $q_c=\sqrt{2mE}/\hbar$, the numerator in Eq.~(\ref%
{eq:cp2}) vanishes linearly and thus $c(q)$ is well-defined for all values
of $q$. To find the behavior of $c(q)$ at large $q$, one can again replace $%
\psi_0(r)$ with the small $r$ approximation. A direct integration gives 
\begin{equation}
c(q)\approx -\frac{8Aa}{q^5a_{\text{ho}}^6\sqrt{1+q^2a^2}},
\end{equation}
revealing a much faster decay of $c(q)$ at large $q$ with respect to the
case when the interaction is turned off simultaneously with the release of
the trap. In particular $c(q)\approx -8Aaa_{\text{ho}}^{-6}/q^5$ if $a_{%
\text{ho}}^{-1}\ll q\ll a^{-1}$ and $c(q)\approx {-8Aa_{\text{ho}}^{-6}/q^6}$
if $q\gg a^{-1}$~\cite{footnote2}.

\begin{figure}[t!]
\includegraphics[width=7.5cm]{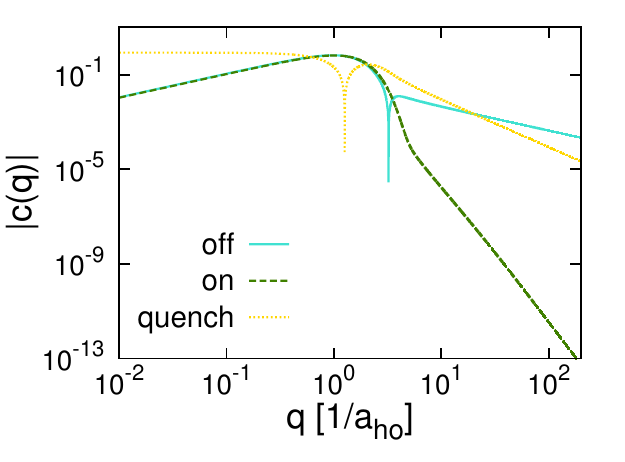}
\caption{Plot of $|c(q)|$ in the absence of interaction
(solid blue line), in the presence of interaction (dashed green line) and
after quenching the scattering length to infinity (dotted orange line). 
Results are shown in log-log scale plot. The absolute value is
added because $c(q)$ changes sign at the dips of the blue and orange lines.
The explicit $q$ dependence of $c(q)$ at small and large $q$ is explained in
the main text.}
\label{fig:cp}
\end{figure}

Figure~\ref{fig:cp} shows the full numerical calculation of $c(q)$ with the
exact initial wave function [Eq.~(\ref{eq:eigenfunction})]. At large momenta
(i.e, in the region of $q\gg a_{\text{ho}}^{-1}$), we find $c(q)\propto 1/q$
in the absence of interaction, whereas it behaves like $1/q^5$ (for $q\ll
a^{-1}$) and $1/q^6$ (for $q\gg a^{-1}$) in the presence of interaction,
with the corresponding coefficients in excellent agreement with the above
analytic estimate. As already pointed out, the momentum distribution
preserves its initial form if the interaction is turned off, whereas it
dramatically changes during the expansion if the interaction is present.
Such a different expansion dynamics is directly connected with the different
behaviors exhibited by $c(q)$ at large momenta.

With the help of $c(q)$, the wave function during the expansion can be, in
fact, calculated straightforwardly: 
\begin{equation}
\psi (r,\tau )=a_{\text{ho}}\int_{0}^{\infty }R_{q}(r)c(q)\exp (-iq^{2}a_{%
\text{ho}}^{2}\tau /2)\frac{dq}{2\pi },  \label{fullwavefunction}
\end{equation}%
and the wave function $\phi (k,\tau )$ in momentum space can be obtained
after a direct Fourier transformation.

\begin{figure}[t!]
\includegraphics[width=7.5cm]{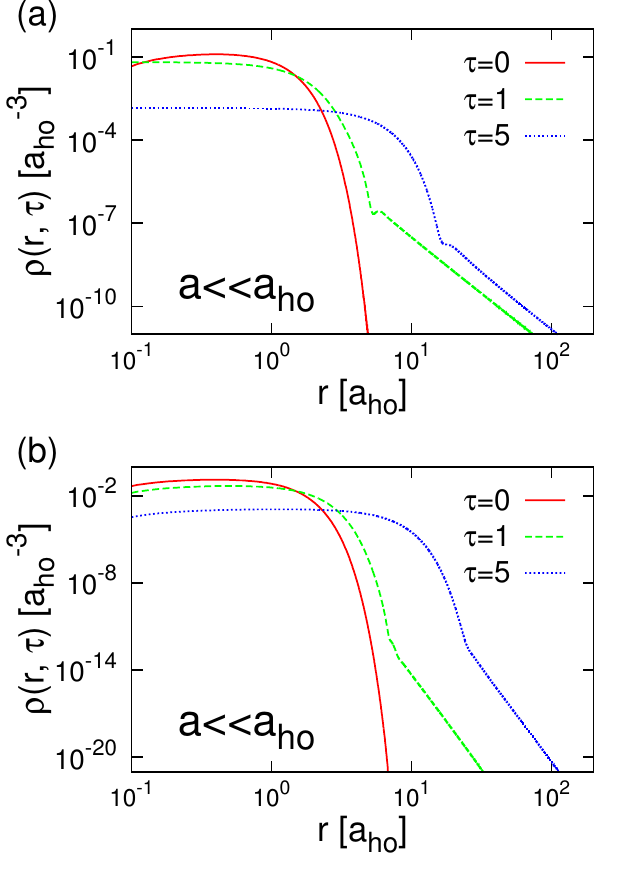}
\caption{Plot of density distribution (a) in the absence and
(b) in the presence of interaction in the expansion. Different lines are
the distributions at different expansion time $\protect\tau=0$ (solid red), $1$
(dashed green), and $5$ (dotted blue). Results are shown in log-log scale plot. The
density distribution $\protect\rho(r, \protect\tau)$ exhibits a $1/r^4$
behavior at large $r$ if the interaction is turned off, whereas it exhibits
a $1/r^{12}$ tail followed by a tail of $1/r^{14}$ if the interaction is
present.}
\label{fig:nr}
\end{figure}

Let us first consider the case where the interaction is turned off just
before the expansion. Since in this case the initial $1/k^4$ large momentum
tail is preserved during the expansion, a $1/r^4$ tail at large distance is
predicted to develop in the density distribution during the expansion. As
shown in Fig.~\ref{fig:nr}(a), a tail of the form $1/r^4$ does actually
appear in $\rho(r, \tau)$ at $r/a_{\text{ho}}\gg \tau$, beyond the
exponential decay. In terms of $c(k)$, the wave function in momentum space
takes the form 
\begin{equation}
\phi(k, \tau)=\sqrt{\frac{1}{2\pi}}\frac{a_{\text{ho}}^{1/2}}{k}c(k)e^{-i%
(k^2a_{\text{ho}}/2) \tau}.
\end{equation}
Since $c(k)\approx 2Aaa_{\text{ho}}^{-2}/k$ at large $k$, thus the large
momentum distribution $n(k, \tau)= \mathcal{C}_0/k^4$ which, as expected,
coincides with the initial one.

If the interaction is not switched off, the expansion dynamics is
dramatically different. In Fig.~\ref{fig:nr}(b) we show the density
distributions resulting from the direct calculation of Eq.~(\ref%
{fullwavefunction}). Due to the presence of interaction, the real space
density always satisfies $\rho(r=a, \tau)=0$. Beyond the exponential decay,
we find that a tail of the form $\rho(r, \tau)\propto 1/r^{12}$ followed by
a tail of the form $\rho(r, \tau)\propto 1/r^{14}$ appears soon after a
short expansion time and is preserved afterwards. This can be understood by
applying to Eq.~(\ref{fullwavefunction}) the saddle-point approximation
which is valid at large $\tau$. At the saddle point $q_s=r/(a_{\text{ho}%
}^2\tau)$, we find 
\begin{equation}
\rho(r, \tau) \approx \frac{1}{2\pi \tau}\frac{|c(q_s)|^2}{r^2a_{\text{ho}}}=%
\frac{32A^2a^2a_{\text{ho}}^7}{\pi}\frac{\tau^{9}}{r^{12}[1+r^2a^2/(a_{\text{%
ho}}^4\tau^2)]}.  \label{eq:rhort}
\end{equation}

\begin{figure}[t!]
\includegraphics[width=7.5cm]{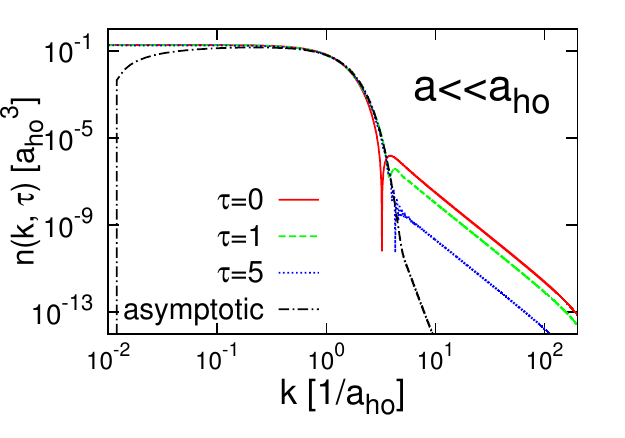}
\caption{Plot of momentum distribution in the presence of
interaction in the expansion. Different lines are the distributions at
different expansion time $\protect\tau=0$ (solid red), $1$ (dashed green), and $5$ (dotted blue) and
the asymptotic momentum distribution (dot-dashed black). Results are shown in
log-log scale plot. At large $k$, the momentum distribution exhibits a tail
of $1/k^4$, whereas the asymptotic momentum distribution exhibits a $1/k^{12}$
tail followed by a $1/k^{14}$ tail.}
\label{fig:nk}
\end{figure}

The absence of the $1/r^4$ tail in the density distribution indicates that
the $1/k^4$ large momentum tail must disappear after a long enough expansion
time. Indeed, as shown in Fig.~\ref{fig:nk}, the large momentum distribution 
$n(k, \tau)\propto 1/k^4$ decreases and shrinks during the expansion,
similarly to the case of the expansion in the presence of unitary
interaction (see Fig.~\ref{fig:unitarity}). Since the momentum distribution
reaches the ballistic regime at a different rate for different momenta,
strong interference oscillations appear in the intermediate momentum regime
connecting the low and large momentum sectors. A careful analysis shows that
the average of these oscillations results in a $1/k^{12}$ behavior with the
increase of time. Therefore, when the ballistic relation is satisfied, the
momentum distribution should exhibit a $1/k^{12}$ tail followed by a $%
1/k^{14}$ tail at $\tau\rightarrow\infty$, consistently with the large $r$
behavior of the density distribution [see Eq.~(\ref{eq:rhort})].

To investigate the time dependence of the $1/k^{4}$ large momentum
tail at large but finite $\tau $, we should calculate the evolution of the
wave function at small $r.$ In this case, only values of $c(q)$ at small $q\lesssim 1/a_{%
\mathrm{ho}}\tau ^{1/2}\sim 1/R(t)$, where $R(t)$ is the radius of the expanded cloud, are important for the calculation of the integral in Eq.~(\ref%
{fullwavefunction}). For these values of $q$ one finds
  $c(q)\approx -A\sqrt{2\pi }a_{\text{ho}}q$ [see Eq.~(\ref{eq:cp}%
)]. Taking into account the asymptotic expression for $R_{q}(r)$ at small $r$,
we find that in this regime the wave function is given by $\psi (r,\tau )\approx
Aa_{\text{ho}}^{-3/2}(a/r-1)/(i\tau )^{3/2}$ and the momentum distribution
behaves like $n(k,\tau )\approx n_{0}(k)/\tau ^{3}$ at large $k$. In
conclusion, the contact decreases as $\mathcal{C}(t)=\mathcal{C}_{0}/\left(
\omega _{\text{ho}}t\right) ^{3}$ for $t\gg 1/\omega _{\text{ho}}$, which
agrees with the numerical calculation of $n(k,\tau )$ shown in Fig.~\ref%
{fig:nk}. The time dependence of the decay is the same as in the case of
the weakly interacting Bose gas discussed in the first part of the paper,
thereby providing a justification of the adiabatic ansatz employed to derive
the $1/t^{3}$ decay law.

As anticipated above, we conclude this section by calculating the probability
that, after release of the trapping potential, the system, instead of
expanding, will occupy a bound state, corresponding to the energy
$E_{B}=-\hbar^2/2ma^2$. This probability is given by $w_{B}=\left\vert
c_{B}\right\vert ^{2},$ where the coefficient $c_{B}$ is given by Eq. (\ref%
{eq:cp2}) with $R_{q}$ replaced by the bound state (dimer) wave function $\psi
_{B}=r^{-1}e^{-r/a}/\sqrt{2\pi a}$. A simple calculation gives $w_{B}=(8/\pi
^{5/2})\left( a/a_{\text{ho}}\right) ^{9}\ll 1$. In the rare case where the
bound state is created, the spatial density will decay exponentially $%
\propto r^{-2}e^{-2r/a}$ and 
the momentum distribution will exhibit the $1/k^4$ behavior for $ka\gg 1$.
However, such a behavior will not affect the measurable density distribution at large distances.

\subsection{Quench dynamics}

In the last part of the work we consider the time evolution of the contact
after a quench of the scattering length from a small value to $a=\infty$
just before the expansion. This problem was already investigated both experimentally~\cite{Makotyn2014,Fletcher2016} and theoretically~\cite{Sykes2014} in the presence of a harmonic trap. The authors of Ref.~\cite{Corson2016} considered the quench dynamics in momentum space of a pair of atoms initially bound by an interaction with $a>0$. In this section, we will consider the dynamics
of a pair of atoms by switching off the trap simultaneously with the quench of the
scattering length. Particularly, we will focus on the investigation of the density distribution of the system
after a large expansion time.

At unitarity, the continuum eigenstates are $R_q(r)=-2a_{\text{ho}%
}^{-1/2}\cos(qr)/r$. Using the approximated form of the initial wave
function $\psi_0(r)\approx Aa_{\text{ho}}^{-3/2}(a_0/r-1)$ at small $r$,
where $a_0$ is the scattering length before the quench, we get $c(q)\approx
-2A/(q^2a_{\text{ho}}^2)$ at large $q$ (see Fig.~\ref{fig:cp}). Applying the
saddle-point approximation to Eq.~(\ref{fullwavefunction}) at the saddle
point $q_s=r/(a_{\text{ho}}^2\tau)$, one then finds 
\begin{equation}
\rho(r, \tau)\approx \frac{1}{2\pi \tau}\frac{|c(q_s)|^2}{r^2a_{\text{ho}}}=%
\frac{2A^2a_{\text{ho}}^3}{\pi}\frac{\tau^3}{r^6}.
\end{equation}

\begin{figure}[t!]
\includegraphics[width=7.5cm]{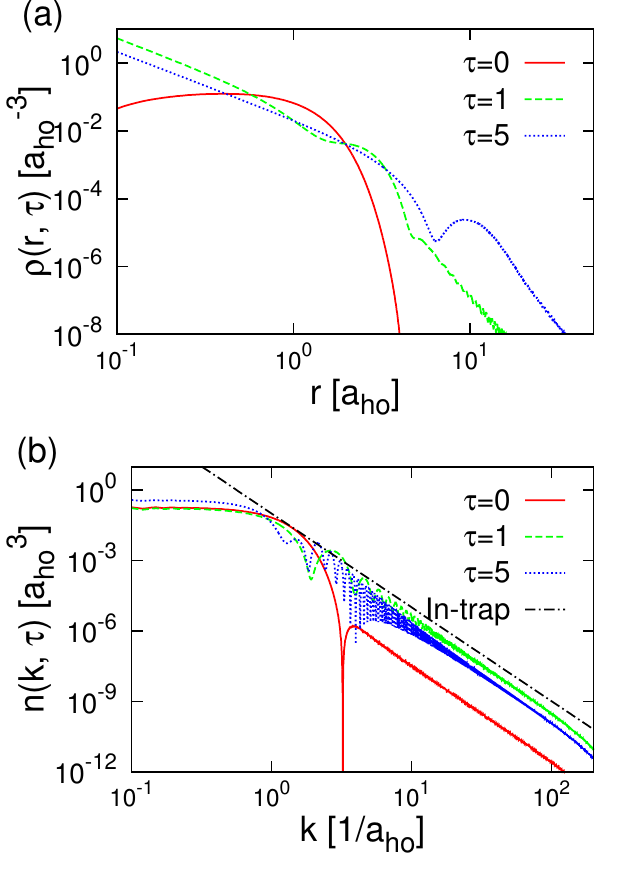}
\caption{Plot of (a) density and (b) momentum distributions
in the expansion with the scattering quenched to $a=\infty$ at $\protect\tau%
=+0$. Different lines are the distributions at different expansion time $%
\protect\tau=0$ (solid red), $1$ (dashed green), and $5$ (dotted blue). The dot-dashed black line
indicates the position of the large momentum tail for the in-trap
equilibrium system with unitarty interaction. Results are shown in
log-log scale plot. At large expansion time, the density distribution
exhibits a tail of $1/r^6$ at large $r$, whereas the momentum distribution
exhibits a tail of $1/k^4$ at large $k$ which first increases at short time
and then decreases.}
\label{fig:quench}
\end{figure}

Using the ballistic relation [Eq.~(\ref{eq:ballistic})], we find that the
asymptotic momentum distribution behaves like $1/k^6$. Consequently, the $%
1/k^4$ tail of the momentum distribution should evolve into $1/k^6$ after
long expansion time in agreement with Ref.~\cite{Corson2016}. Figure~\ref{fig:quench} shows the time-dependent
evolution of the density and momentum distributions in the expansion after
quenching the scattering length from $a_0=0.05a_{\text{ho}}$ to $a=\infty$.
A tail of the form $1/r^6$ is developed in the density distribution, in good
agreement with the above analytical calculation. The momentum distribution
can be separated into two stages. At shorter times, the two-body
correlations increase due to the quench and the $1/k^4$ tail increases and
quickly reaches the in-trap equilibrium value, as investigated in Ref.~\cite%
{Sykes2014}. At longer times, the expansion dynamics dominates and the
contact decreases, in agreement with our previous results for the unitary
interaction. Since the rate of the evolution is slow, a new $1/k^6$ tail is
expected to appear after a long expansion time.

It is finally worth investigating the effect of the quenching of the
scattering length to a finite value $a_{f}$ satisfying the condition $%
\left\vert a_{f}\right\vert \ll $ $a_{\text{ho}}$. In this case the system,
after expansion, will exhibit the density shape
\begin{equation}
\rho(r, \tau)\approx
\frac{2A^2}{\pi a_\text{ho}}\frac{\tau}{r^4}\frac{(a_f-a)^2}{(1+r^2a_f^2/a_\text{ho}^4\tau^2)}
\end{equation}
holding for $r/a_\text{ho}\gg \tau$. According to Eq.~(\ref{eq:ballistic}), the corresponding
asymptotic momentum distribution takes the form
\begin{equation}
n_\text{asym}(k)=\frac{2A^2}{\pi a_\text{ho}^3}\frac{(a_f-a)^2}{k^4}
\label{eq:quenchasym}
\end{equation}
in the interval $a_\text{ho}^{-1}\ll k \ll |a_{f}|^{-1}$ where, if $a_f >0$,  we can safely
ignore the contribution caused by the formation of a bound state (dimer) during the expansion~\cite{footnote3}.

\section{Conclusion}

\label{sec:conclusion} 
In this paper we have investigated the
time dependence of the density and momentum distribution of various
expanding physical systems after the release from a harmonic trap, with
special focus on the behavior of the contact parameter. We have explicitly
shown that the momentum distribution changes dramatically if the interaction
is present during the expansion and the large momentum tail decreases and
eventually disappears for large times. On the other hand, differently from
the case when the interaction is switched off before the expansion, the
density distribution, which is the observable quantity in the time-of-flight
experiments, does not exhibit any $1/r^4$ behavior, even at intermediate
times where the $1/k^4$ tail in the momentum distribution is still visible.
So far, our investigations have been devoted to isotropically trapped
systems. The presence of anisotropy in the trapping potential is not
expected to introduce major differences in the large $k$ behavior of the
momentum distribution of the expanding gas, although its role might deserve
further theoretical investigation.

On the basis of our results we conclude that the $1/r^4$ density tail,
observed after the release of an anisotropic trap in the recent experiment
by Chang, \textit{et al.} carried out on a weakly interacting Bose gas of
metastable $^4$He atoms~\cite{Chang2016}, is still elusive and remains an
open problem to understand.

\textit{Note added.} Recently, we noticed a
preprint by Gharashi and Blume~\cite{Gharashi2016} on the evolution
of the contact for a smoothly released one-dimensional harmonic trap.

\begin{acknowledgments}
We would like to thank David Clem\'ent and Alain Aspect for stimulating our interest in the problem of the contact during the expansion and  for informing us about their recent experiment on the expansion of metastable $^4$He atoms. Useful correspondence and discussions with Shina Tan, Stefano Giorgini, and Christophe Salomon are acknowledged. C.Q. thanks Yangqian Yan for helpful discussions. This work was supported by the QUIC grant of the Horizon2020 FET program and by Provincia Autonoma di Trento.
\end{acknowledgments}

\begin{appendix}
\section{One-body density matrix}
\label{appdenx1}
Using the scaling law [Eq.~(\ref{castin})], an exact relationship can be obtained also for the time dependence of the one-body density matrix,
\begin{equation}
\rho^{(1)}(\mathbf{r},\mathbf{r}^\prime, t) = \frac{1}{b^3} e^{-i(m/2\hbar)(\dot{b}/b)(\mathbf{r}^2-{\mathbf{r}^\prime}^2)}\rho_0^{(1)}\left(\frac{\mathbf{r}}{b},\frac{\mathbf{r}^\prime}{b}\right).
\label{eq:onebodyrho}
\end{equation}
Introducing the sum and difference of these two position vectors, $\mathbf{R}=(\mathbf{r}+\mathbf{r}^{\prime})/2$ and $\mathbf{s}=(\mathbf{r}-\mathbf{r}^{\prime})$, and taking the Fourier transformation, one then obtains the exact  result
\begin{equation}
n(\mathbf{k},t)= \frac{1}{(2\pi)^3}\frac{1}{b^3}\int d\mathbf{R}d\mathbf{s}\left[ e^{-i\mathbf{s}\cdot\left[
\mathbf{k}-(m\dot{b}/\hbar b)\mathbf{R}
\right]} \rho_0^{(1)}\left(\frac{\mathbf{R}}{b},\frac{\mathbf{s}}{b}\right)\right]
\label{eq:onebodynk}
\end{equation}
for the momentum distribution at time $t$, in terms of the initial value of the one-body density matrix.

To calculate the asymptotic momentum distribution $n_{\text{asym}}=\lim_{t\to\infty}n(\mathbf{k}, t)$ it is convenient to introduce the rescaled variable $\tilde{R}=R/b(t)$.  At large times, where $b(t) \to \omega_{\text{ho}}t$, one then easily finds the result
\begin{eqnarray}
&&\lim_{t\to\infty}n(\mathbf{k}, t) \nonumber \\
&=&\left( \frac{\hbar}{m\omega_\text{ho}}\right)^3 \int d\tilde{\mathbf{R}}\delta
(\tilde{\mathbf{R}}- \frac{\hbar}{m\omega_\text{ho}}\mathbf{k})
\rho_0^{(1)}\left(\tilde{\mathbf{R}},0\right)
 \nonumber \\
&=&
\left(\frac{\hbar}{m\omega_{\text{ho}}} \right)^3\rho_0\left(\tilde{\mathbf{R}}=\frac{\hbar}{m\omega_{\text{ho}}}\mathbf{k} \right),
\end{eqnarray}
which coincides with the result [Eq.~(\ref{eq:asym_unitary})] derived in Sec.~\ref{sec:manybody}  employing the ballistic relation.

The time dependence of the contact at finite expansion times can  be also obtained by calculating the asymptotic behavior of Eq.~(\ref{eq:onebodynk}) at large $k$. For large momenta satisfying $k\gg \frac{m\dot{b}}{\hbar b}R_0$ which implies $ka_{\text{ho}}\gg R_0/a_{\text{ho}}$, with $R_0$ the initial size of the trapped gas, the variable $\mathbf{R}$ can be ignored in the exponential factor of Eq.~(\ref{eq:onebodynk}) and the momentum distribution approaches  the scaling behavior
\begin{equation}
n(\mathbf{k}, t) \to b^3(t) n[\mathbf{k}b(t)].
\end{equation}
As a consequence, the contact takes the form 
\begin{equation}
\mathcal{C}(t)=\frac{\mathcal{C}_0}{b(t)},
\end{equation}
which coincides with the result  obtained using the adiabatic ansatz for the unitary Fermi gas.

\end{appendix}


\begin{thebibliography}{99}
\bibitem{footnote2} It is worth noting that the regime $c(q)\propto 1/q^6$
does not exist in the hard sphere model where the range of interaction $%
r_0=a $, because the whole theory is valid at the condition $q r_0\ll 1$.
However, this regime can be reached if $a\gg r_0$.
\bibitem{footnote3} For larger value of $k$, satisfying the condition $ka_f\gg 1$,
the momentum distribution of the system is
instead dominated by the formation of the dimer and
is given by the same Eq.~(\ref{eq:quenchasym}), with an additional multiplying factor
4. However, as discussed at the end of Sec.~\ref{sec:twobody}(B), the
dimer will not affect the behavior of the the expanding density distribution at large distances.
\end{thebibliography}
\end{document}